\documentclass[12pt,tightenlines,
    onecolumn,
    preprintnumbers,
    amsmath,
    amssymb,
    prd,
    showpacs,
    showkeys
]{revtex4}
\usepackage{times}
\usepackage{colordvi}
\input colordvi
\usepackage{graphicx}%
\usepackage{amsmath}
\usepackage{amssymb}
\usepackage{bm}
\usepackage{enumerate}
\usepackage{color}

\newcommand{\nn}{\nonumber}

\newcommand{\beq}{\begin{equation}}
\newcommand{\eeq}{\end{equation}}
\newcommand{\beqa}{\begin{eqnarray}}
\newcommand{\eeqa}{\end{eqnarray}}

\newcommand{\bg}{\textbf{g}}

\newcommand{\bra}[1]{\langle #1|}
\newcommand{\ket}[1]{|#1\rangle}

\newcommand{\cF}{\mathcal{F}}

\newcommand{\cU}{\mathcal{U}}

\newcommand{\tr}{\triangleright}
\newcommand{\poin}{Poincar\'{e} }
\begin{document}
\title{Braided Statistics from Abelian Twist in $\kappa$-Minkowski Spacetime}

\author{Hyeong-Chan Kim}
\email{hckim@cjnu.ac.kr}
\affiliation{Division of Liberal Arts, Chungju National University,
Chungju 380-702, Republic of Korea}
\author{Youngone Lee}
\email{ youngone@daejin.ac.kr}
\affiliation{Department of Physics, Daejin University, Pocheon, Gyeonggi 487-711, Korea}
\author{Chaiho Rim}
\email{rimpine@sogang.ac.kr}
\affiliation{Department of Physics and Center for Quantum Spacetime, 
Sogang University, Seoul 1221-742, Korea}
\begin{abstract}
\bigskip
$\kappa$-deformed commutation relation between quantum
operators is constructed via abelian twist deformation 
in $\kappa$-Minkowski spacetime.
The commutation relation is written in terms of universal 
$R$-matrix satisfying braided statistics. 
The equal-time commutator function turns out to vanish 
in this framework. 
 
\end{abstract}
\pacs{11.10.Nx, 04.60.Ds, 02.20.Uw}
\keywords{$\kappa$-Minkowski spacetime, twist deformation,
braided statistics, commutator function}
\maketitle

\section{Introduction}

It is widely believed that conventional concept of spacetime
should be changed in Plank scale. Noncommutative spacetime
is a way to obtain a deformed symmetry in the Planck scale~\cite{dopl}.
In particular, $\kappa$-Minkowski spacetime ($\kappa$-MST)~\cite{majid}
is represented in terms of the commutation relation
(with  $i, j $ spatial index $1,2,3$ ) 
\begin{eqnarray}\label{kappa}
[x^0, x^i]= \frac{i}{\kappa} x^i, \quad [x^i,x^j]=0\,,\nn
\end{eqnarray}
which has the merit of rotational symmetry in space.
$\kappa$-MST is originated from
$\kappa$-deformed Poincar\'{e} algebra~\cite{lukierski}
and is realized in many ways~\cite{luki2,Wess2,Nowak,mel,Bu2}.
The deformed Poincar\'{e} algebra allows two invariant parameters 
and is noted to be related with doubly special relativity~\cite{dsr}.
However, it is realized that the field theoretical approach 
in this spacetime is afflicted with difficulties, such as
violation of causality~\cite{bb}, 
instability of vacuum structure~\cite{casimir} 
and energy momentum non-conservation 
in interacting field theory~\cite{luki}. 

To understand $\kappa$-MST more deeply, twist formalism 
such as Jordanian  twist~\cite{jordantwist,kulish,Borowiec},  
$\kappa$-like deformation of quantum Weyl and conformal algebra
~\cite{Ballest} and the light-cone $\kappa$-deformation
of Poincar\'e algebra~\cite{lightlike}
has been tried. 
In the following, abelian twist 
in~\cite{Bu:2006dm,Kim:2008mp,Kim:2009jk} will be used.
The abelian twist is based on 
inhomogeneous general linear group in 4-dimensions, 
in which the Poincar\'e group is embedded.

The purpose of this paper is to construct a deformed product of
quantum field operators in the abelian twist formalism. 
There have been  attempts to understand the deformed 
statistics, using an intertwiner which commutes 
with $\kappa$-\poin transformation
~\cite{Govindarajan:2008qa,Arzano:2008bt,Young:2007ag}.
A new product has been proposed for 
$\kappa$-deformed quantum field operators
\cite{Daszkiewicz:2007ru}
which preserve the bosonic statistics 
of the $n$-particle states.
Our twist approach is endowed with universal $R$-matrix
~\cite{twist-hopf}
and can provide a way to understand the statistics of many particle states in $\kappa$-MST. 

The order of the paper is the following.
The abelian twist of field operators
is introduced in Sec.~\ref{sec:module}. 
The deformed product (called $\star$-product) 
between momentum operators 
is defined in Sec.~\ref{sec:starproduct}. 
This leads to the $\star$-product between operators defined 
at different spacetime points. 
Deformed commutation relations 
are obtained in terms of universal $R$ matrix
in Sec.~\ref{sec:comm}. 
Sec.~\ref{Remark} is the summary and discussion.

\section{$igl(4,R)$ and momentum operators} \label{sec:module}

There are 20 generators $Y=\{M^a_{~b}\,,  P_a \}$
($a,b =0,1,2,3$) in $\cU(\bg\backsimeq igl(4,R))$.
The generators act on a coordinate module as 
$ M^a_{~b}\tr x^c =i x^a \delta_b^c$ and 
$P_a \tr x^b = i \delta_a^b$.
Thus, $Y$ is represented in coordinate basis as
\begin{eqnarray}
(M^a_{~b} \tr  \phi)(x)
     &=& i x^a \frac{\partial}{\partial x^b} \phi(x)\,, \\
(P_a \tr \phi)(x)
    &=& i \frac{\partial}{\partial x^a} \phi(x) \,. \nn
\end{eqnarray}
It is, however, desirable to allow $Y$ to act on (covariant) momentum operator as well 
since quantum mechanical operators are conveniently 
used in momentum space. 

Suppose one Fourier-transforms  $\varphi(x)$:
\begin{equation}
{\varphi}(x)=\int_p ~e^{-ip\cdot x}~ \phi_p 
\end{equation}
where $p\cdot x =p_a x^a$
and the upper and lower indices are
to be distinguished.  
$\int_p$~ represents $igl(4,R)$-invariant measure,
which is achieved by introducing a metric~$g_{ab}$
(signature + - - -), ~
$\int_p = \int
\frac{d^4p_a} {(2\pi)^4 \sqrt{-g} }= \int
\frac{d^4p^a\,\sqrt{-g} } {(2\pi)^4  }$.
The raising and lowering of the indices are given 
using the metric, $ p_a = g_{ab}~p^b $. 
~$g $ is the determinant of $g_{ab}$.
The metric transforms under 
{\it global homogeneous} 
$igl(4,R)$ transformation $M_{ab}$
and is independent of 
coordinate or momentum so that 
$x\cdot x = x^a x^b g_{ab}$,
$p\cdot x = p^a x^b g_{ab}$ and 
$p^2 = p^a p^b  g_{ab}$ 
are invariant.

Through the Fourier-transformation 
one can define action of $Y$ on momentum function 
$\phi_p$:
\beqa
(P_a \tr\phi)(x)
&=&\int_p~
 \left( i \frac{\partial}{\partial x^a}  e^{-ip\cdot x}\right) ~ \phi_p
  \equiv \int_p~
e^{-ip\cdot x}~ \left(P_{a}\tr \phi_p \right)  
\\
(M^a_{~b} \tr\phi)(x)
&=&\int_p~
 \left( i x^a \frac{\partial}{\partial x^b}  
 e^{-ip\cdot x}\right) ~ \phi_p
  \equiv \int_p~
e^{-ip\cdot x}~ \left(M^a_{~b}\tr \phi_p \right)
\,.
\eeqa
Explicitly, one has
\beq
\label{action-p}
P_a \tr \phi_p
=p_a \,\phi_p  \,, \quad
e^{i\alpha D}\tr \phi_p = e^{3 \alpha}
\phi_{(p_0,e^\alpha {\bf p})}  \,,
\eeq
where $D =\sum_{i=1}^3 M^i_{~i}$ is the space dilatation, 
${\bf p}=(p_1,p_2,p_3)$ is the space momentum covariant-vector.
Here we use the fact 
$e^{i\alpha  D} (x^i) e^{-i \alpha D}
=  (x^{i})^{-\alpha}$ and ${\bf p} \mapsto e^{\alpha}{\bf p}$.
Note that the momentum ${\bf p}$ 
in Eq.~(\ref{action-p}) can be redefined 
up to the rigid $igl(4,R)$ transformation. 
So, we assume that a specific reference frame is chosen 
so that  $\phi_p$ transforms according to 
Eq.~(\ref{action-p}).

To proceed, let us 
divide the field operator into
positive and negative frequency parts: 
\beqa
\phi_p &=&
[\theta(p_0) \phi^{(+)}_p+\theta(-p_0)\phi^{(-)}_{p}]\,.
\eeqa
The action of $Y$ on $\phi^{(m)}_p$ ($m=+,-$) is 
given by 
$ P_a\tr \phi^{(m )}_p = m\, p_a \phi^{(m)}_p $ 
and 
$ e^{i\alpha D}\tr\phi^{(m)}_p
 = e^{3 \alpha} \phi^{(m)}_{(p_0,e^\alpha {\bf p})}$.
It is noted that this decomposition is possible because 
time-like domain remains time-like 
under the global $igl$(4,R) transformation  
because the transformation of the metric $g_{ab}$ 
guarantees 
$p^2= p^a p^b g_{ab}$ invariant.  
In the following, 
we will use a block diagonal basis of the metric, 
$g_{00}=1, g_{0i}=g_{i0}=0$. 
If $g_{0i}\neq 0$, 
one may use Anorwitt-Deser-Misner decomposition.

\section{$\star$- Product}
\label{sec:starproduct}
In Ref.~\cite{Bu:2006dm}, abelian twist element is presented:
\begin{eqnarray}
\label{kelement}
{\cal F_\kappa}=
\exp\left[\frac{i}{2\kappa}\left(E\otimes D-D\otimes E\right)\right],
\end{eqnarray}
where $E(=P_0)$ commutes with $D$. 
The co-product of twisted  Hopf algebra is defined as
\beq
\label{co-product}
\Delta_\kappa(Y)=\cF \cdot \Delta Y \cdot \cF^{-1}
=\sum_{i}Y_{(1)i}\otimes Y_{(2)i}\,.
\eeq
Using  $\kappa$-deformed product 
$a \otimes_\kappa b \equiv \cF^{-1}\tr a\otimes b$,
one defines the $\star$-product of momentum operators
\cite{Bu:2006ha},
\beq \label{star}
(a\star b)
= \circ~(a\otimes_\kappa  b)\,
\eeq
where $\circ$ is the ordinary product of operators.
The $\star$-product of $\phi^{(\pm)}_p$ and $\phi^{(\pm)}_q$
are written explicitly;
\begin{equation}
\label{star_mn}
\phi^{(m)}_p\star \phi^{(n)}_q
= e^{\frac{3(-m\, p_0+ n\, q_0)}{2\kappa} }
\phi^{(m)}_{(p_0,e^{n\, q_0/{2\kappa}}{\bf p})}\circ
\phi^{(n)}_{(q_0,e^{ -m\, p_0/{2\kappa}}{\bf q})}\,.
\end{equation}

As the consequence, one can define $\star$-product between 
operators at two different spacetime points. 
Consider $\star$-product of two functions 
with positive frequency part:
\begin{equation} 
\label{xy-product}
 \varphi ^{(+)} (x) \star_{xy}  \varphi^{(+)} (y)
 =  \left( \int_{p_+} \, e^{-ip \cdot x} \, \phi^{(+)}_p \right) 
 \star_{xy}  \left( \int_{q_+} \, e^{-iq \cdot y} \,  \phi^{(+)}_q\right)
\end{equation}  
where $ \star_{xy}$ emphasizes that $\star$-product 
is evaluated between two different spacetime points 
and $\int_{p_+}$ denotes $\int_p$   with $p_0 \ge 0$. 
If one evaluates the $\star$-product
Eq.~(\ref{xy-product}) 
before integration,
two different evaluations  should be consistent:  
\beq
\int_{p_+} \int_{q_+} \,   
\left( e^{-ip \cdot x} \star_{xy}  e^{-iq \cdot y} \right)
~\left( \phi^{(+)}_p \circ \ \phi^{(+)}_q  \right)
=\int_{p_+} \int_{q_+} \,  
 \left(e^{-ip \cdot x} \circ  e^{-iq \cdot y} \right)
~\left( \phi^{(+)}_p \star \phi^{(+)}_q \right)\,.
\eeq  
This consideration leads to the $\star_{xy}$-product 
\begin{equation}
e^{-ip \cdot x} \star_{xy} e^{-iq \cdot y} 
= e^{-i (p_0 x^0+q_0 y^0 +  e^{-q_0/2\kappa}~p_i x^i
    + e^{p_0/2\kappa}~ q_i y^i)} \,.
\end{equation}  
Note that similar relation holds 
for positive and/or negative frequency functions 
and therefore, the positivity condition 
for $p_0$ and $q_0$ can be lifted.
This result can be considered as a natural generalization of Eq.~(\ref{star}) on coordinate space. 
The same formula of $\star_{xy}$-product 
appears in \cite{Daszkiewicz:2007ru}.

\section{$\star$- Commutation relation}
\label{sec:comm}

The twisted Hopf algebra naturally 
induces a braided statistics as shown in \cite{twist-hopf}
\begin{equation}
\circ\left[ \phi^{(m)}_p \otimes_\kappa \phi^{(m)}_q \right] -
\circ\left[R^{-1}\tr \phi^{(m)}_q\otimes_\kappa\phi^{(m)}_p\right]=0 \,.
\end{equation}
where  $R\equiv\cF_{21}\cF^{-1}= \cF^{-2}$ is the universal $R$-matrix and satisfies the Yang-Baxter equation.
Explicitly,  
\begin{equation}
\circ\left[R^{-1}\tr\phi^{(m)}_q \otimes_\kappa \phi^{(m)}_p\right] 
= e^{ \frac{m(p_0-q_0)}{\kappa}}
\phi^{(m)}_{(q_0,e^{ m\, p_0/{\kappa}}{\bf q})} \star\phi^{(m)}_{(p_0,e^{ -m\, q_0/{\kappa}}{\bf q})} 
\end{equation}
and one has the deformed commutation relations
\begin{equation}
\phi^{(m)}_p\star \phi^{(m)}_q -
e^{ \frac{m(p_0-q_0)}{\kappa}}
\phi^{(m)}_{(q_0 , e^{m\,p_0/\kappa}{\bf q})} 
\star\phi^{(m)}_{(p_0 , e^{-m\, q_0/\kappa}{\bf p} )}
=0 \,.
\label{deformed}
\end{equation}
As $( p_0 ~\mathrm{and} ~q_0 ) \ll \kappa  $,
this relation reduces to the 
ordinary commutation relation, 
$  \phi^{(m)}_p\circ \phi^{(m)}_q = 
\phi^{(m)}_q\circ \phi^{(m)}_p $. 

Likewise, one may have the commutation relation between 
positive and negative frequency parts,
\begin{equation} 
\label{+-star}
 \phi^{(+)}_p \star \phi^{(-)}_q  
- e^{\frac{p_0+q_0}{\kappa}} 
\phi^{(-)}_{(q_0,e^{p_0/{\kappa}}{\bf q})}\star
\phi^{(+)}_{(p_0,e^{ q_0/{\kappa}}{\bf p})}
= f(p, q) 
\end{equation}
since 
\beq
\circ\left[R^{-1}~\tr \phi^{(-)}_q\otimes_\kappa \phi^{(+)}_p\right] =   
e^{\frac{p_0+q_0}{\kappa}} 
\phi^{(-)}_{(q_0,e^{p_0/{\kappa}}{\bf p})}\star
\phi^{(+)}_{(p_0,e^{q_0/{\kappa}}{\bf q})}\,.
\eeq
$f(p, q) $ is to be determined.  
Suppose the vacuum is annihilated 
by the positive frequency part: 
$ \phi_p ^{(+)} \left| \mathrm {vac}  \right\rangle =0 $. 
(In this way, the vacuum defines 
the signature of positive frequency).
Then the vacuum expectation value  
\[
\Phi_2(x,y) =
\left\langle \varphi (x) \star_{xy} \varphi(y)  \right\rangle
= 
\left\langle \varphi^{(+)} (x) \star_{xy} \varphi^{(-)}(y)  \right\rangle
\]
can be written as 
\beqa
\nn
\Phi_2(x,y)&=& \int_{p_+} \int_{q_+}\,
e^{-i p \cdot x} \star_{xy} e^{i q \cdot y} \,
\bra{ \mathrm {vac} }
 \phi_p^{(+)} \circ \phi_q^{(-)} 
\ket{\mathrm {vac} }
\\
&=& \int_{p_+} \int_{q_+}\,
e^{-i p \cdot x}  e^{i q \cdot y} \,
\bra{ \mathrm {vac} }
\phi_p^{(+)} \star \phi_q^{(-)} 
\ket{\mathrm {vac} }\,.
\eeqa 
Typically, 
$\bra{ \mathrm {vac} }\phi_p^{(+)} 
\circ \phi_q^{(-)}  \ket{\mathrm {vac} }$ 
is given on-shell,
\[
\left\langle \phi_p^{(+)} 
\circ \phi_q^{(-)}  \right\rangle
=\sqrt {-g}~  2 p_0 ~(2\pi)^5  \delta ({\bf p}-{\bf q})
~\delta(p^2 -m^2) ~\delta(q^2 -m^2) 
~\theta(p_0)~ \theta(q_0)  \,.
\]
This will determine $f(p,q)$
from Eqs.~(\ref{star_mn}) and (\ref{+-star});
\beqa
f(p,q) 
\label{fpq}
&=& \left\langle \phi_p^{(+)} 
\star \phi_q^{(-)}  \right\rangle 
\nn\\
&=& \sqrt {-g}~ 
e^{\frac{-3(p_0+  q_0)}{2\kappa}}~ 
{2 p_0 } ~ (2\pi)^5   
\delta ({\bf p} _q -{\bf q} _p) 
 ~\delta(p_q^2 -m^2) ~\delta(q_p^2 -m^2) 
~\theta(p_0)~ \theta(q_0)
\eeqa
where ${\bf p} _q ={\bf p} e^{-q_0/2\kappa}$, 
${\bf q}_p ={\bf q} e^{-p_0/2\kappa}$,
$p_q^2 = p_0p^0 + p_ip^ie^{-q_0/\kappa} $ 
and 
$q_p^2 = q_0 q^0 + q_iq^ie^{-p_0/\kappa} $. 

\section{Summary and discussions} \label{Remark}

In this report, we investigate $\star$-commutation relation 
in $\kappa$-Minkowski spacetime by using abelian twist. 
We start from the undeformed bosonic commutation relation
and arrive at deformed relations~(\ref{deformed}) and 
(\ref{+-star}) in terms of universal $R$ matrix.

The twist formalism on momentum space 
allows the $\star_{xy}$-product 
between operators at different points. 
Thus, one may evaluate the $\star$- commutator function:
\beqa
\label{commfunction}
&& \langle \phi (x)\star \phi(y)- \phi(y)\star \phi (x) \rangle \nn \\
&&~~~~~~~~
=
 \int_{p_+} \int_{ q_+}\,
 \left[ e^{-i p\cdot x} \star_{xy}  e^{i q\cdot y}
-e^{-ip\cdot y} \star_{yx}  e^{i q\cdot x}
\right]
\bra{\mathrm {vac}} \phi^{(+)}_p\circ \phi^{(-)}_q\ket {\mathrm {vac}} \nn\\
&&~~~~~~~~
=\int_{p_+} 2\pi
~\delta(p^2 -m^2)
~\left(
e^{-i p_0\cdot (x^0-y^0)-i p_i (x^i -y^i)e^{p_0/2\kappa}}
-(x  \leftrightarrow y )
\right)\,.
\eeqa
The star commutator function
deviates  from that of the commutative one.
Still, it vanishes at equal time $x^0=y^0$ since 
the on-shell condition is symmetric in $p_i$.
This conclusion holds for space-like region 
because $(x-y)^a \star (x-y)_a = (x-y)^a (x-y)_a$
is rigid $igl(4,R)$-invariant. 

In this abelian twist approach, 
if one quantizes an on-shell field, 
then one will face a trouble in realization of 
$R$-matrix. On-shell condition fixes the energy 
in terms of the spatial momentum 
but the $R$-matrix or the twist 
${\cal F_\kappa}$ behaves as if 
$p_0$ and ${\bf p}$ are independent.
Hence, under the on-shell condition  
$D\otimes E$ and $E \otimes D $ 
do not commute each other.
This is the reason why 
the commutation relation 
Eq.~(\ref{+-star}) is evaluated 
after the introduction of the vacuum.

It is noted that positive and negative frequency 
part is fixed relative to the vacuum.  On the other, 
the signature of $p_0$ is frame-dependent. 
This implies that the vacuum may be frame-dependent
or time flow is to be defined accordingly.  
In addition, one needs proper understanding of 
the role of Poincar\'e symmetry  
in this abelian twist field theory context. 
These questions are to be dealt with 
carefully in the future publication. 

Finally, one tempts to allow $p$-dependent metric transformation 
so that the scale factor 
in Eq.~(\ref{action-p}) is absent and 
on-shell condition 
is not changed in Eq.(\ref{fpq}). 
This transformation will result in non-trivial 
non-Riemannian geometry
which might be relevant 
if one considers non-trivial manifold 
such as Finsler manifold.

\begin{acknowledgments}
This work was supported by the Korea Science and Engineering Foundation (KOSEF) grant funded by the Korea Government(MEST)(R01-2008-000-21026-0 (R)) and through the Center for Quantum Spacetime(CQUeST) of Sogang University with grant number R11-2005-021(R) and  by the Korea Research Foundation grant funded by the Korea Government(MOEHRD), KRF-2008-314-C00063(H.-C. K. and Y. L.).
\end{acknowledgments} 

\vspace{4cm}

\end{document}